\documentclass{article}

\usepackage[english]{babel}
\usepackage{booktabs}

\usepackage[letterpaper,top=2cm,bottom=2cm,left=3cm,right=3cm,marginparwidth=1.75cm]{geometry}

\usepackage{amsmath}
\usepackage{graphicx}
\usepackage[colorlinks=true, allcolors=blue]{hyperref}
\usepackage{main-defs}
\usepackage{subcaption}
\title{Analysis of Hardware Synthesis Strategies for Machine Learning in Collider Trigger and Data Acquisition}
\author{
     Haoyi Jia\textsuperscript{1,2}, Abhilasha Dave\textsuperscript{2}, Julia Gonski\textsuperscript{2}, Ryan Herbst\textsuperscript{2}\\
    \textsuperscript{1}Stanford University, Stanford, CA, USA \\ 
    \textsuperscript{2}SLAC National Accelerator Laboratory, Menlo Park, CA, USA \\
}

\begin{document}
\maketitle
\tableofcontents

\begin{abstract}
To fully exploit the physics potential of current and future high energy particle colliders, machine learning (ML) can be implemented in detector electronics for intelligent data processing and acquisition. 
The implementation of ML in real-time at colliders requires very low latencies that are unachievable with a software-based approach, requiring optimization and synthesis of ML algorithms for deployment on hardware. 
An analysis of neural network inference efficiency is presented, focusing on the application of collider trigger algorithms in field programmable gate arrays (FPGAs). 
Trade-offs are evaluated between two frameworks, the SLAC Neural Network Library (SNL) and \hls, in terms of resources and latency for different model sizes.  
Results highlight the strengths and limitations of each approach, offering valuable insights for optimizing real-time neural network deployments at colliders.
This work aims to guide researchers and engineers in selecting the most suitable hardware and software configurations for real-time, resource-constrained environments.
\end{abstract}


\clearpage

\section{Introduction}
\label{sec:intro}

Potential for precision measurement of ultra-fast and small-scale fundamental processes dictates the specifications of physics experiments and their readout systems. 
Driven by these scientific goals, a variety of next-generation facility concepts share a common feature of unprecedented data rates and dataset sizes. 
Combined with unique experimental features such as ultra-fast latencies, high radiation doses, cryogenic temperatures, and spatial constraints, the electronic readout systems of these detectors are often subject to extensive R\&D to ensure state-of-the-art performance. 

In collider physics, data is produced when the initial particle collision produces byproducts which pass through sensitive detector material. 
Front-end data acquisition (DAQ) systems at the detector sensor sample the detector cells at the particle collision rate, perform some preliminary data processing at-source, and transmit the data off-detector for further processing and eventual storage. 
At the Large Hadron Collider (LHC), the on-detector electronics must sample at 40 Mhz and are subject to high radiation doses, together requiring implementation on application specific integrated circuits (ASICs). 
To manage the overwhelming $\mathcal{O}$(1) Tb/s data rate, an intermediate stage between the on-detector readout and off-detector storage known as a ``trigger" evaluates each collision event within 25 ns to determine if it should be written to disk or discarded.
The low latency of trigger operation, along with the need for configurability to allow for varying trigger algorithm designs, motivate the use of  field programmable gate arrays (FPGAs) for the first trigger stage. 

While the challenges of collecting and triggering data at today's scientific facilities already require creative design and cutting-edge technology, they are only expected to be exacerbated at future machines.
The High Luminosity LHC will deliver collisions with an expected number of simultaneous interactions of approximately 200, over three times the current scenario in Run 3 of the LHC, resulting in a substantial increase in the detector occupancy and number of final state particles.
The Future Circular Collider (FCC), an international facility proposed to be hosted at CERN, comprises two stages: an $e^+e^-$ collider for precision measurements, followed by a $pp$ collider at $\sqrt{s} =$ 100 TeV for novel discovery potential which will deliver an expected exascale data rate~\cite{Benedikt:2651300}. 
Similar challenges of real-time data rate management are present in photon science; the upgrade of the Linac Coherent Light Source (LCLS) X-ray free electron laser at SLAC National Accelerator Laboratory must collect and process data from light pulses with 1 MHz repetition rate, for an estimated $\mathcal{O}$(100) Gb/s rate. 
Such high data rate experiments often also have an increased complexity of the underlying physical process, which must be well-modeled to inform the subsequent DAQ stages. 

Recent progress with the implementation of machine learning (ML) into hardware platforms such as FPGAs can address the challenges of high data rate experiments by enabling intelligence in the trigger and/or DAQ systems of next-generation physics experiments~\cite{10.3389/fdata.2022.787421, duarte2022fastmlsciencebenchmarksaccelerating}. 
The implementation of an ML algorithm into a hardware platform requires the use of high-level synthesis (HLS) to convert an algorithm developed in software with standard ML libraries into a hardware description language. 
This process results in a hardware version of an ML algorithm that can be subsequently configured to an FPGA or used in an ASIC design. 

To achieve optimized network latency and efficient resource allocation in FPGA-based neural networks, it is critical to consider the data input rate and the impact of HLS on hardware implementation, considering in particular the following factors: 

\paragraph{Data Input Rate and Latency Implications}
The rate and manner in which data arrives for processing are fundamental in determining the achievable latency in FPGA-based neural networks. If data arrives as a single, sequential stream, the network processing speed is inherently constrained by this input rate. 
Even with extensive parallelism and advanced architectures, the system’s maximum performance cannot surpass the rate at which data is fed into it. 

On the other hand, by having multiple data streams or input batches processed simultaneously, the system can exploit parallel computation more effectively, significantly impacting the overall latency and improving throughput.
For example, suppose a neural network is designed to process data in a pipelined fashion. 
In that case, a continuous and parallelized data inflow enables various stages of computation to operate concurrently, which minimizes idle time and optimizes latency. This approach not only reduces waiting periods between data arrivals but also aligns well with FPGA architectures, where resources can be reconfigured to manage several operations in parallel.

\paragraph{High-Level Synthesis and Variability in FPGA Implementations}

The hardware implementations generated by HLS tools exhibit variability in latency and resource usage for identical algorithms based on application-specific optimization strategies. 
The choice of synthesis strategy-- whether aimed at reducing latency, minimizing resource usage, or balancing both-- directly influences the FPGA’s efficiency and suitability for different applications. For instance, in fields like high energy physics, where data processing often involves real-time analysis of vast data volumes from complex events, FPGA-based solutions need to be customized with a synthesis strategy that meets rigorous latency and resource constraints.
In HLS, the same neural network algorithm can be synthesized with varying parameters, resulting in implementations with distinct operational characteristics. 
A latency-focused synthesis may prioritize faster processing by utilizing more FPGA resources, while a resource-optimized synthesis may trade-off some latency for a more economical use of FPGA components. These trade-offs must align with the application’s needs—especially in high-stakes environments like physics experiments, where even microsecond-level latencies could impact results.

\paragraph{Managing Complexity in Model Size and Architecture}
The increasing complexity of machine learning models used in FPGA-based applications introduces further challenges for efficient hardware implementation. Large neural networks or sophisticated architectures, such as deep convolutional or recurrent layers, demand substantial FPGA resources and can introduce latency bottlenecks if not optimized correctly. 
To address these challenges, model compression techniques, quantization, or sparsity exploitation can be integrated during synthesis to reduce model size without significantly impacting performance.
Additionally, HLS tools enable designers to apply fine-grained control over the synthesis process, allowing for resource reuse and pipelining to meet both resource-efficiency and latency requirements. 
These optimizations are particularly beneficial in FPGA designs because they enable an increase in throughput while keeping resource consumption within acceptable limits, even for larger models.
The increased complexity of the underlying physical process or event motivates the use of larger and more sophisticated ML models, making the above features essential for design consideration. 

\textsc{} 

This paper presents an analysis of FPGA resource utilization and latency for ML models implemented using two HLS strategies: the SLAC Neural Network Library (SNL)~\cite{herbst2023implementationframeworkdeployingai} and~\hls~\cite{Duarte:2018ite, fastml_hls4ml}. 
\hls~is an open-source Python package to create firmware deployments of ML models commonly used in high energy physics, with support for a broad set of ML architectures and frameworks. 
SNL is developed with a focus on the LCLS application, targeting high performance, low latency, and dynamically reconfigurable ML implementations. 
The comparative study of these two strategies can offer insight into the trade-offs between power efficiency and performance, particularly in the context of real-time data processing and high-throughput scientific applications. 

A generic collider trigger application is used as a benchmark, specifically considering fully connected variational autoencoders designed for real-time anomaly detection on collision events. 
Several model benchmarks are designed and synthesized, scaling up the number of nodes per layer, to specifically investigate the potential for modern synthesis frameworks to meet the demands of larger-scale, real-time ML models in next-generation facilities.
The resulting synthesis comparison sheds light on the strengths and limitations of each approach, offering valuable guidance for selecting the most suitable framework for specific use cases in the rapidly evolving landscape of edge computing and machine learning acceleration.

\section{Synthesis Frameworks}
\label{sec:synthesizers}

\subsection{SLAC Neural network Library (SNL)}



\begin{figure}
    \centering
    \includegraphics[width=0.7\linewidth]{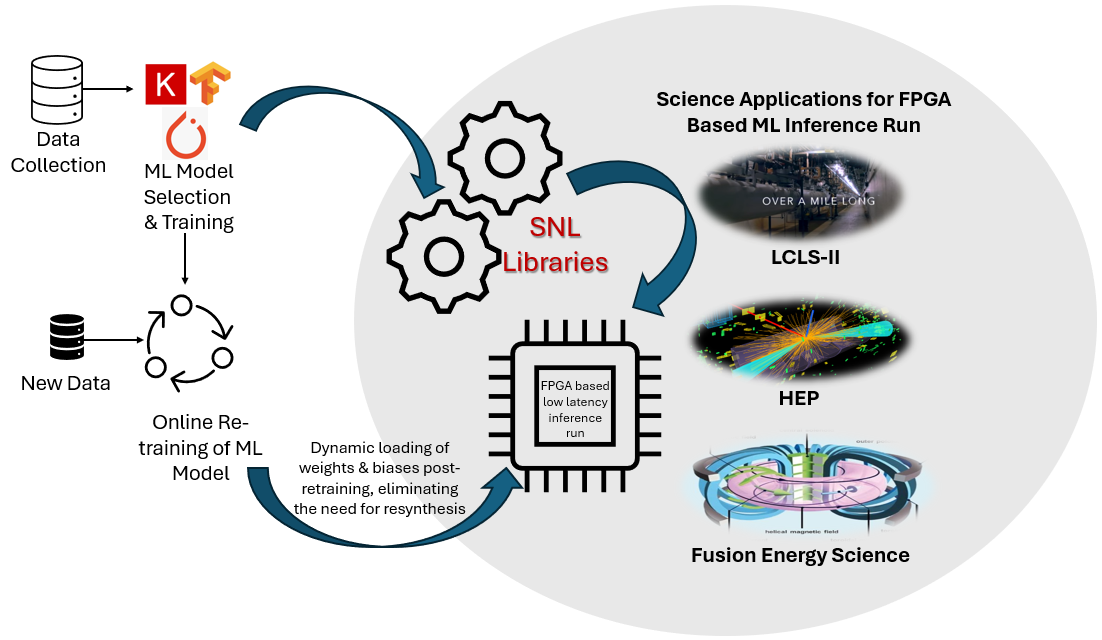}
    \caption{High-level design flow for the SLAC Neural network Library (SNL).
    \label{fig:snl_designflow}}
\end{figure}

The SLAC Neural Network Library (SNL) is a framework for enabling the the deployment of machine learning inference models on FPGAs. 
This approach facilitates the construction of data processing pipelines characterized by ultra-low latency and high throughput. 
These edge computing systems process raw detector outputs by performing essential tasks such as preprocessing, vetoing, and classifying frames before transmitting compressed data downstream for further analysis or storage.
Figure~\ref{fig:snl_designflow} shows the design flow with SNL.  

One of the most significant advantages of SNL is its ability to dynamically load weights and biases. 
This means that once a model has been synthesized and loaded onto the FPGA, retraining the model does not require re-synthesizing the entire setup. 
Instead, new weights and biases can be uploaded dynamically, allowing for immediate inference runs without the time-consuming process of FPGA resynthesis. 
This feature greatly enhances the flexibility and efficiency of using FPGAs in environments where models may need to be frequently updated or fine-tuned.
This approach is not only vital for current detector technologies but also holds tremendous potential for future collider experiments and other high-performance applications where low latency is critical. 
By streamlining the translation of machine learning architectures into optimized FPGA code, SNL provides a robust solution for real-time ML based data processing across a wide range of scientific domains. 

The SNL framework is implemented as a collection of C++ templates optimized for the Xilinx Vitis HLS development environment, a choice that balances performance with ease of use. Leveraging C++ templates in this context enables efficient hardware-accelerated computation while preserving a straightforward, user-friendly design. The templates for ML layers are carefully aligned in naming, ordering, and functionality with the methods used in Python’s Keras library. This intentional mirroring ensures that users familiar with Keras can smoothly transition to SNL’s C++ templates, finding the parameters and layer definitions both intuitive and recognizable. This alignment also allows users to reference Keras’s robust documentation as a supplementary guide when working with the SNL framework.

In addition to the template design, the SNL framework employs a streaming interface within its neural network architecture, enhancing efficiency for latency-sensitive, real-time applications. The streaming interface enables each layer to start processing data as soon as the necessary output from the preceding layer becomes available, which minimizes cumulative latency. This contrasts with a memory interface approach, which introduces delays by requiring each layer to wait until the previous one has fully completed processing. Such delays can be a drawback in time-sensitive tasks, where latency reduction is more critical than throughput.

For applications like triggering mechanisms and feedback loops, where rapid response times are essential, the SNL framework’s streaming interface supports continuous data flow between layers and across network inputs and outputs. By combining the high-performance capabilities of the Xilinx Vitis HLS environment with Keras-inspired template familiarity and an optimized streaming architecture, the SNL framework effectively meets the demands of real-time, hardware-accelerated neural network development.


\subsection{\hls}

\begin{figure}
    \centering
    \includegraphics[width=0.9\linewidth]{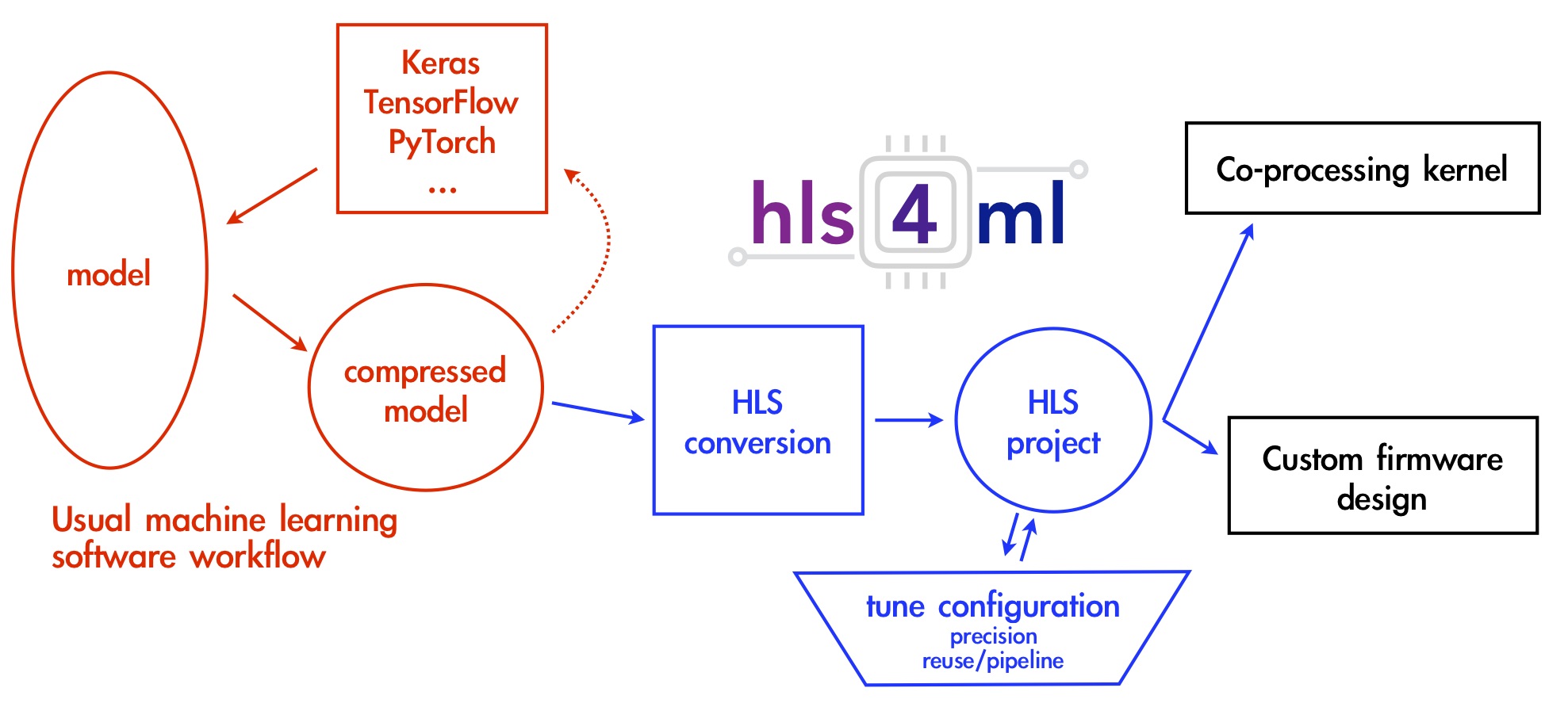}
    \caption{High-level design flow for \hls~\cite{fastml_hls4ml}.
    \label{fig:hls_designflow}}
\end{figure}

\hls~is similarly an innovative framework designed to facilitate the implementation of machine learning models on FPGAs using HLS~\cite{Duarte:2018ite, fastml_hls4ml}.
\hls~streamlines the design process, allowing researchers and engineers to optimize and deploy machine learning models efficiently, thereby bridging the gap between software-based machine learning and hardware acceleration.
Figure~\ref{fig:hls_designflow} shows the design flow with \hls. 

By translating neural network architectures into FPGA-compatible code, \hls~enables rapid prototyping and deployment of low-latency, high-throughput models directly on hardware. 
This approach is particularly beneficial for applications requiring real-time data processing, such as those found in high-energy physics, where the ability to perform complex computations at the edge is crucial. 
\hls~has been used broadly across high energy physics to deploy real-time ML models. 
The application chosen for this benchmarking study, the autoencoder-based trigger algorithm, was implemented using \hls~in the CMS experiment at the Large Hadron Collider~\cite{CMS-DP-2024-059, CMS-DP-2023-079}.

\section{Models \& Dataset}
\label{sec:models}

The synthesis strategy comparison is done for an example task of a real-time anomaly detection algorithm for a collider trigger system. 
The anomaly detection capability is enabled by a variational autoencoder (VAE)~\cite{kingma2022autoencodingvariationalbayes}, which is trained in an unsupervised way to learn the distribution of Standard Model collider phenomena. 

The VAE does this by learning to reconstruct its input after compression to a lower dimensional latent space, minimizing reconstruction error with respect to the truth input values. 
The latent space of a VAE comprises Gaussians described by a mean $\mu$ and standard deviation $\sigma$; the latent space is honed during training such that background-like inputs are encoded similarly in the latent space. 
When an anomalous element is evaluated by the VAE, it will be poorly reconstructed and be separable in the latent space from the background distribution.
In this way, the loss of the VAE upon evaluation can be useful for the task of anomaly detection. 

VAE models are trained over a simulated dataset of proton collision events that emulates a typical data stream collected by an LHC detector, pre-filtered by requiring the presence of at least one electron or muon~\cite{govorkova2021lhcphysicsdatasetunsupervised}.
Each VAE model is trained over 3.2 million simulated collision events for 40 epochs. 
Input events are modeled by a total of 57 values: the four-vector quantities, namely \pt, $\eta$, and $\phi$ of the 10 highest momentum jets, 4 highest momentum muons, and 4 highest momentum electrons produced in the proton collision, along with the \pt~and $\eta$ of the missing transverse energy. 
The VAE consists of a simple 3 layer fully connected multilayer perceptron (MLP) encoder and a symmetric decoder, connected by the latent space. Each layer in the encoder and decoder is a fully connected dense layer followed by a ReLU activation function. 
For the VAE to evaluate within stringent latency constraints, only the encoder is deployed on the FPGA, and an event is determined to be anomalous based on a modified Kullback-Leibler divergence score using only the encoded mean $\mu$ of an event in the Gaussian latent space representation~\cite{Govorkova_2022}. 
Therefore, synthesis is only done for the encoder stage of a VAE.

\begin{figure}
    \centering
    \includegraphics[width=1\linewidth]{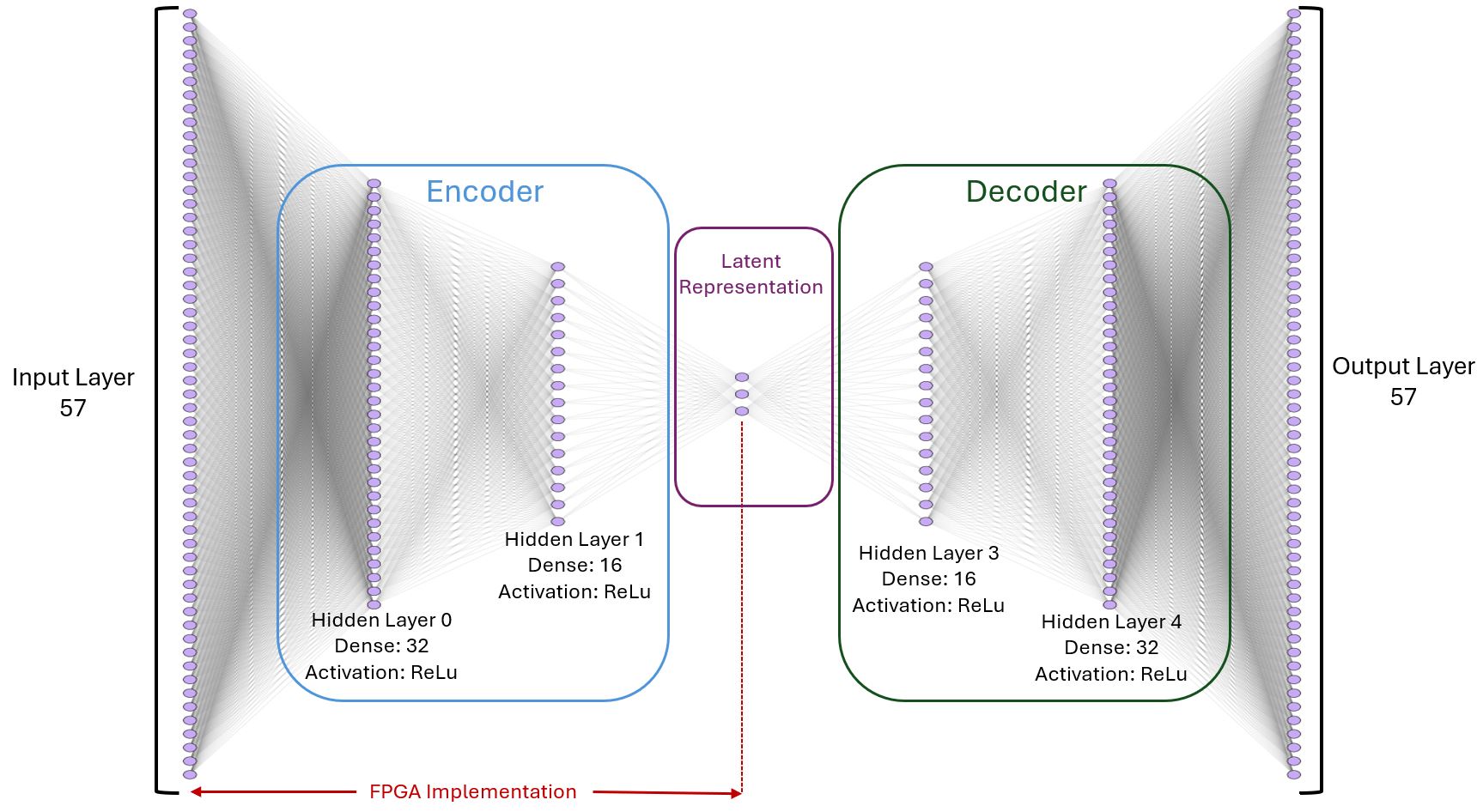}
    \caption{Architecture diagram for the smallest VAE encoder benchmark model (Model 1). Two other models, 2 and 3, share the same structure but have a number of nodes per layer that is a factor of 2 and 4 larger, respectively. Only the encoder stage is implemented on the FPGA, as indicated. 
    \label{fig:models}} 
\end{figure}

The size of the model in terms of trainable parameters impacts the ability to synthesize it for FPGA implementation.
To explore the varying capabilities of the two synthesis strategies, three encoder model benchmark points are used.
Model 1 is built to represent a model size that can deliver good anomaly detection performance with events at the LHC~\cite{Govorkova_2022}, and comprises 3 fully connected layers with dimensions 32, 16, and 3 for a total of 2435 trainable parameters.
Two larger benchmarks, Models 2 and 3, have the same structure, but number of nodes in each layer is doubled and quadrupled respectively, leading to models with 5990 and 16460 trainable parameters. 
Figure~\ref{fig:models} shows a diagram of Model 1, representing the same architecture for Models 2 and 3 with the exception of the number of nodes per layer. 

Table~\ref{tab:model_size} provides a summary of the three benchmark models, their number of nodes per layer, and their total trainable parameters. 
The study of inference efficiency in FPGA implementation for large models with a high number of parameters seeks to inform the development of advanced triggers for collider scenarios with an unprecedented number of overlapping particles and interactions. 

\begin{table}
    \centering
    \caption{Number of nodes in each layers and trainable parameters in each model \label{tab:model_size}}
    \begin{tabular}{l|cccc}
        \toprule
        \textbf{Models} & \textbf{First Layer} & \textbf{Second Layer} & \textbf{Third Layer} & \textbf{Trainable Parameters} \\
        \midrule
        Model \#1 & 32   & 16   & 3  &  2435 \\
        Model \#2 & 64   & 32   & 6  &  5990 \\
        Model \#3 & 128   & 64   & 12  & 16460 \\
        \bottomrule
    \end{tabular}
\end{table}

\section{Synthesis Results}
\label{sec:results}

Results are given as comparisons of resource usage and latency for the FPGA inference runs between SNL and \hls~syntheses. 
The resources considered are those of a modern FPGA, specifically block random access memory (BRAM), digital signal processors (DSPs), flip-flops (FFs), and look-up tables (LUTs). 
All model sizes are capable of achieving a $\mathcal{O}$($\mu$s) runtime upon synthesis, making them viable for operation in current FPGA-based trigger systems. 

Trade-offs between latency and resource usage are possible when translating the ML algorithm to a hardware implementation.
To ensure a comparison of SNL and \hls~with the most similar configurations, the reuse factor is used to produce syntheses with approximately the same latency. 
Therefore, four different synthesis strategies are shown: SNL,  \hls~with a resource strategy, \hls~with a Latency strategy with the reuse factor set to 1 (``optimized latency"), and \hls~with a strategy to achieve a comparable latency to SNL by allowing the resource factor to exceed 1 (``matching latency"). 
A hardware target of Alveo U200 ~\cite{amd_alveo_u200} accelerator card is used with total available resources of 4320 BRAM, 6840 DSPs, 2364480 FFs, and 1182240 LUTs under 200Mhz clock rate. 

To reduce resource usage, model inputs, parameters, and values passing between layers are quantized following training (post-training quantization)~\cite{han2016deepcompressioncompressingdeep}. 
Two different quantization levels are used, \texttt{ap\_fixed<32,16>} and \texttt{ap\_fixed<16,8>}, to best represent the possible applications of the VAE model in realistic trigger scenarios and understand how resources and latency are mapped in varying precision.
For all \texttt{hls4ml} syntheses, data are passing through layers in parallel instead of streaming I/O type, whereas for SNL the data is streamed between layers. 

\begin{figure}
    \centering
    \includegraphics[width=0.45\linewidth]{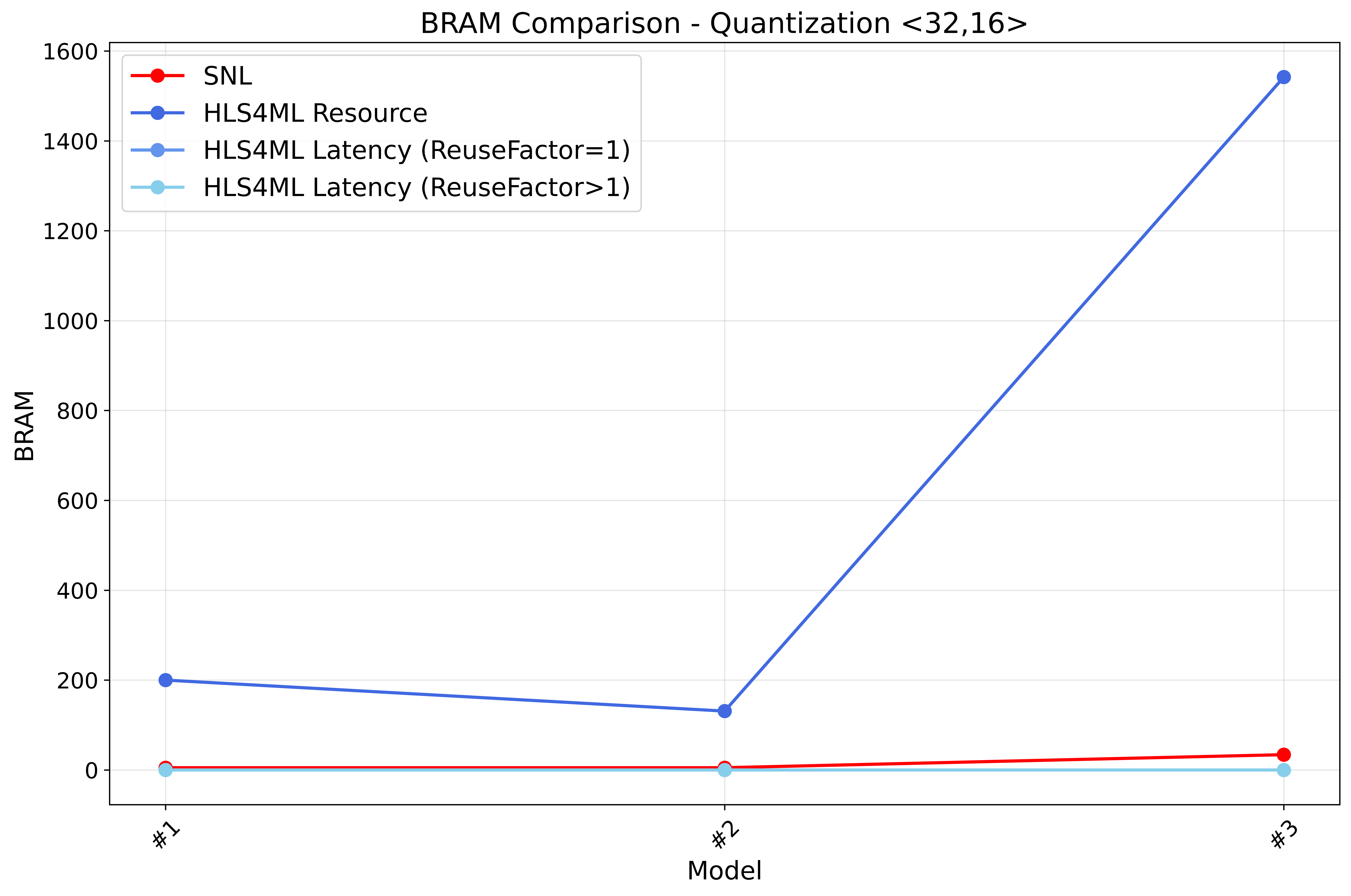}
    \includegraphics[width=0.45\linewidth]{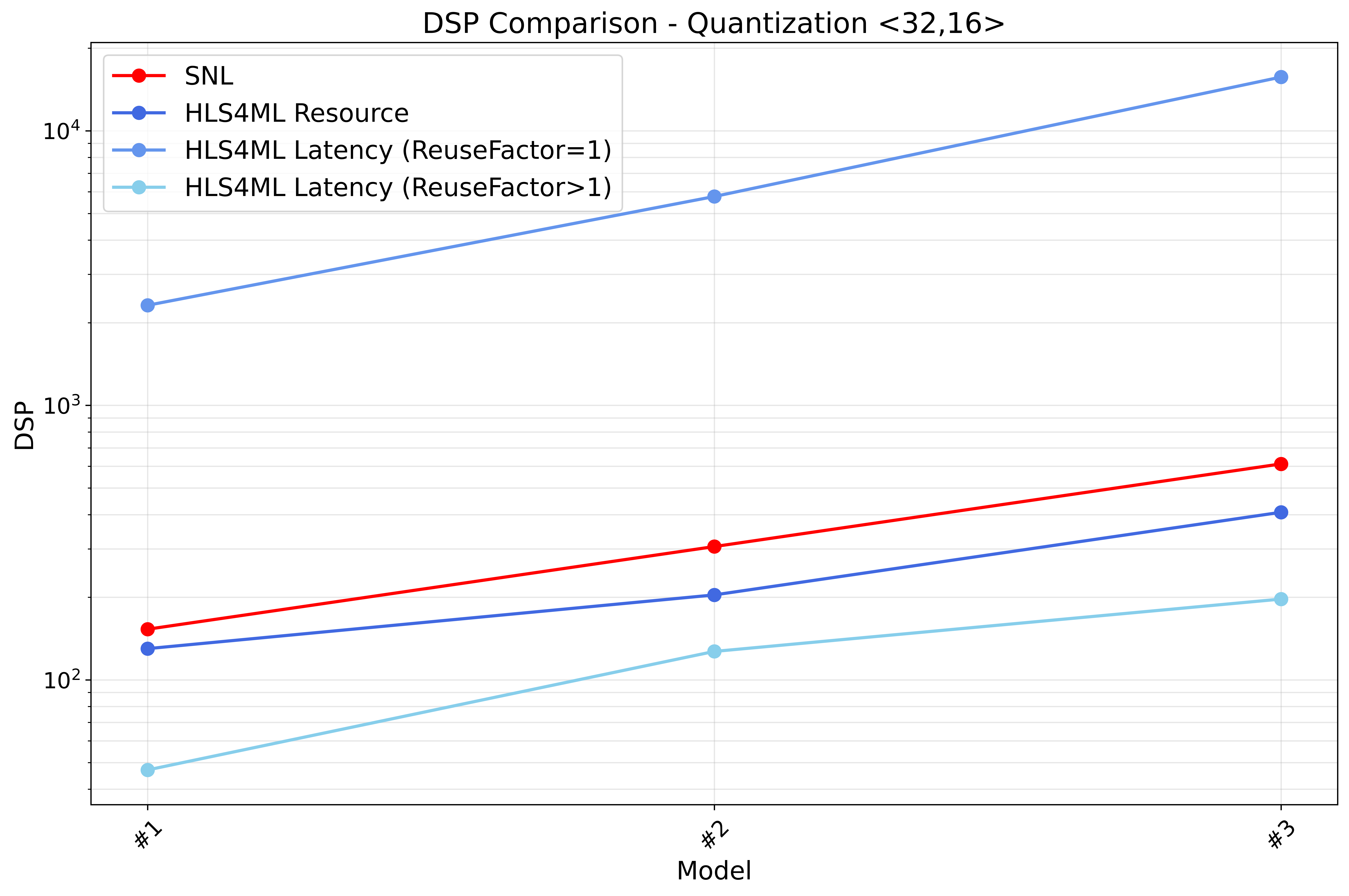}
    \includegraphics[width=0.45\linewidth]{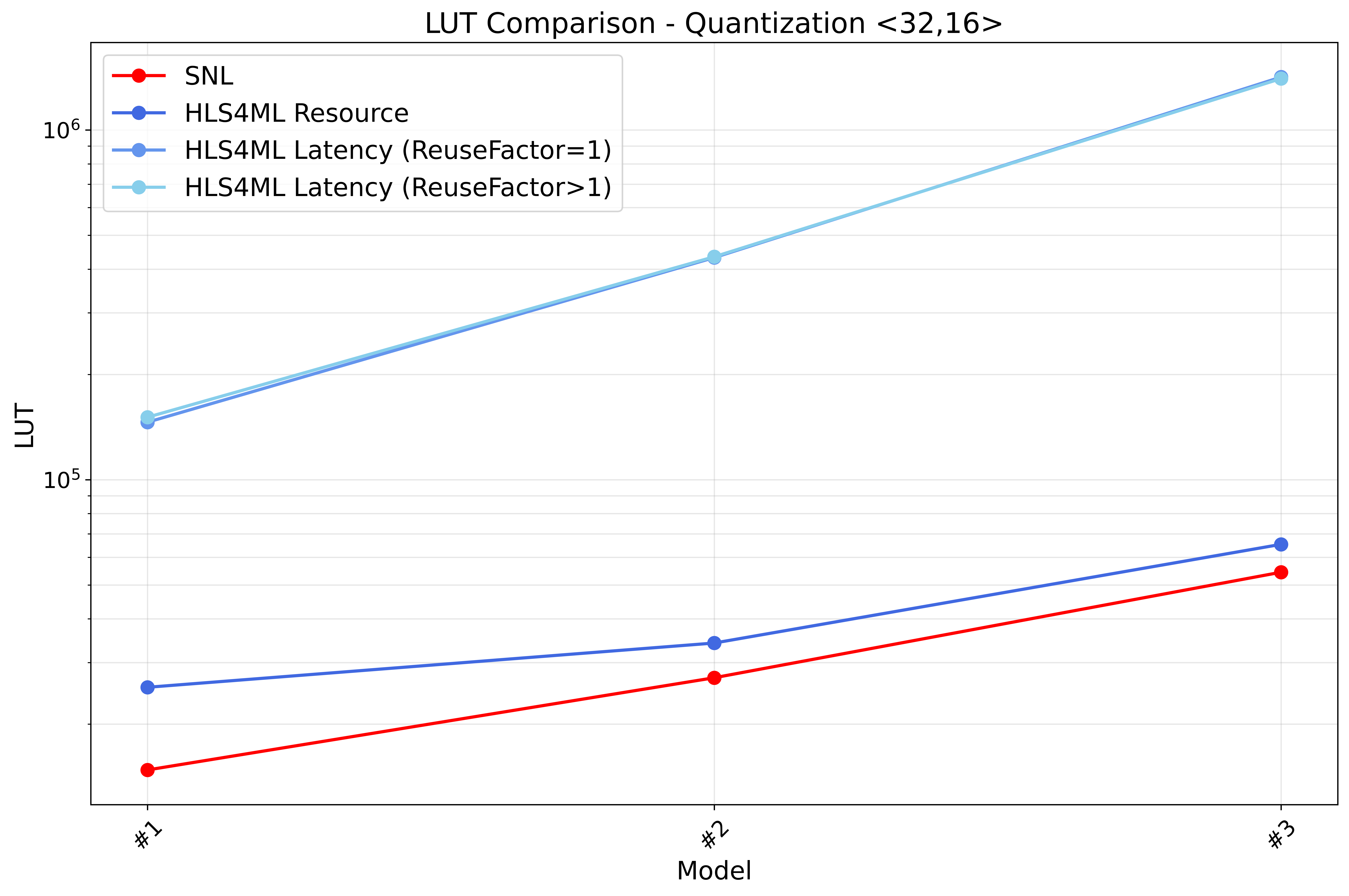}
    \includegraphics[width=0.45\linewidth]{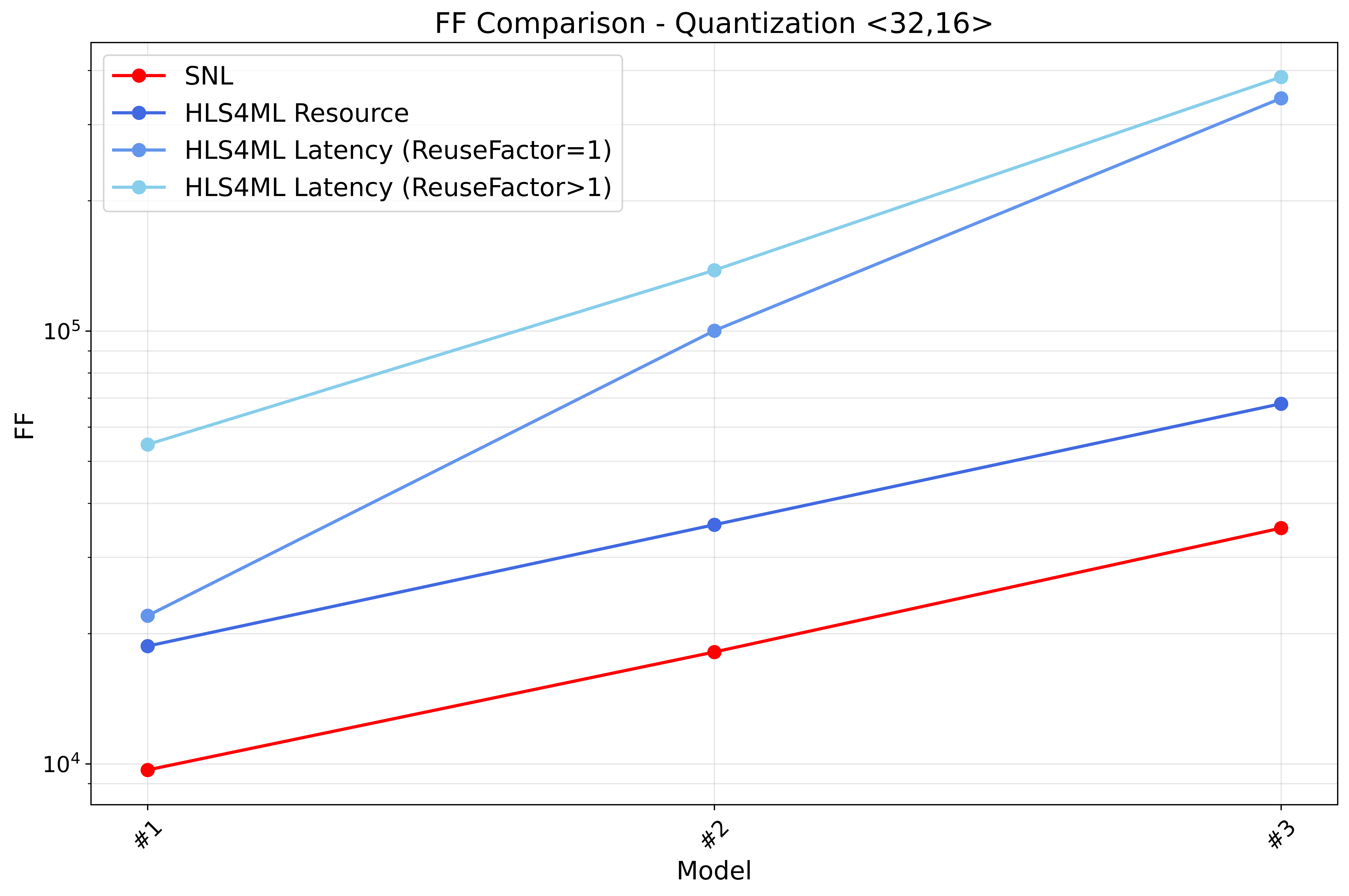}
    \includegraphics[width=0.45\linewidth]{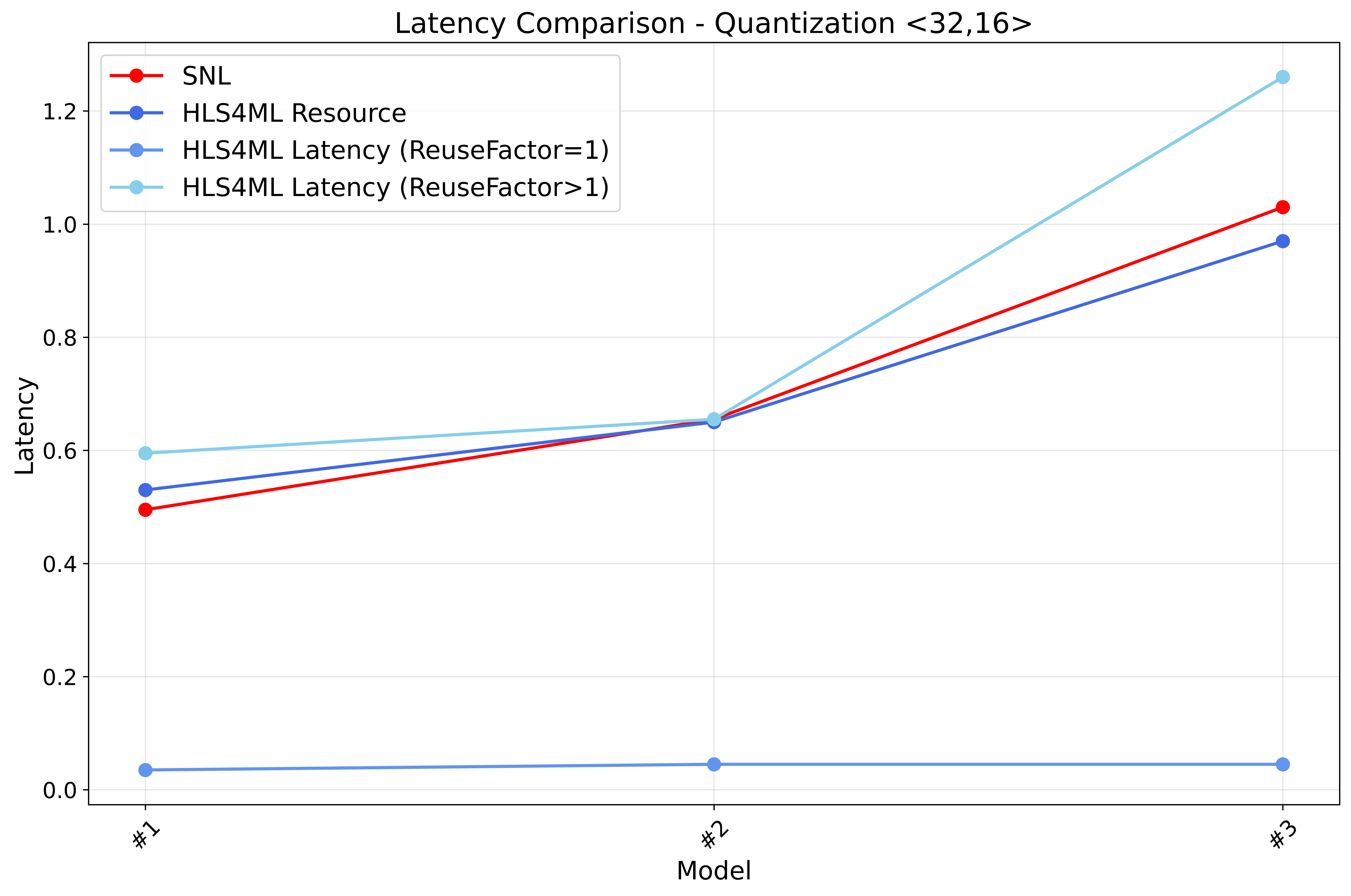}
    \caption{Plots comparing the usage of BRAM, DSPs, LUTs, FFs, and latency for the 3 benchmark models with \texttt{ap\_fixed<32,16>} quantization.
    \label{fig:plots_3216}} 
\end{figure}

\begin{figure}
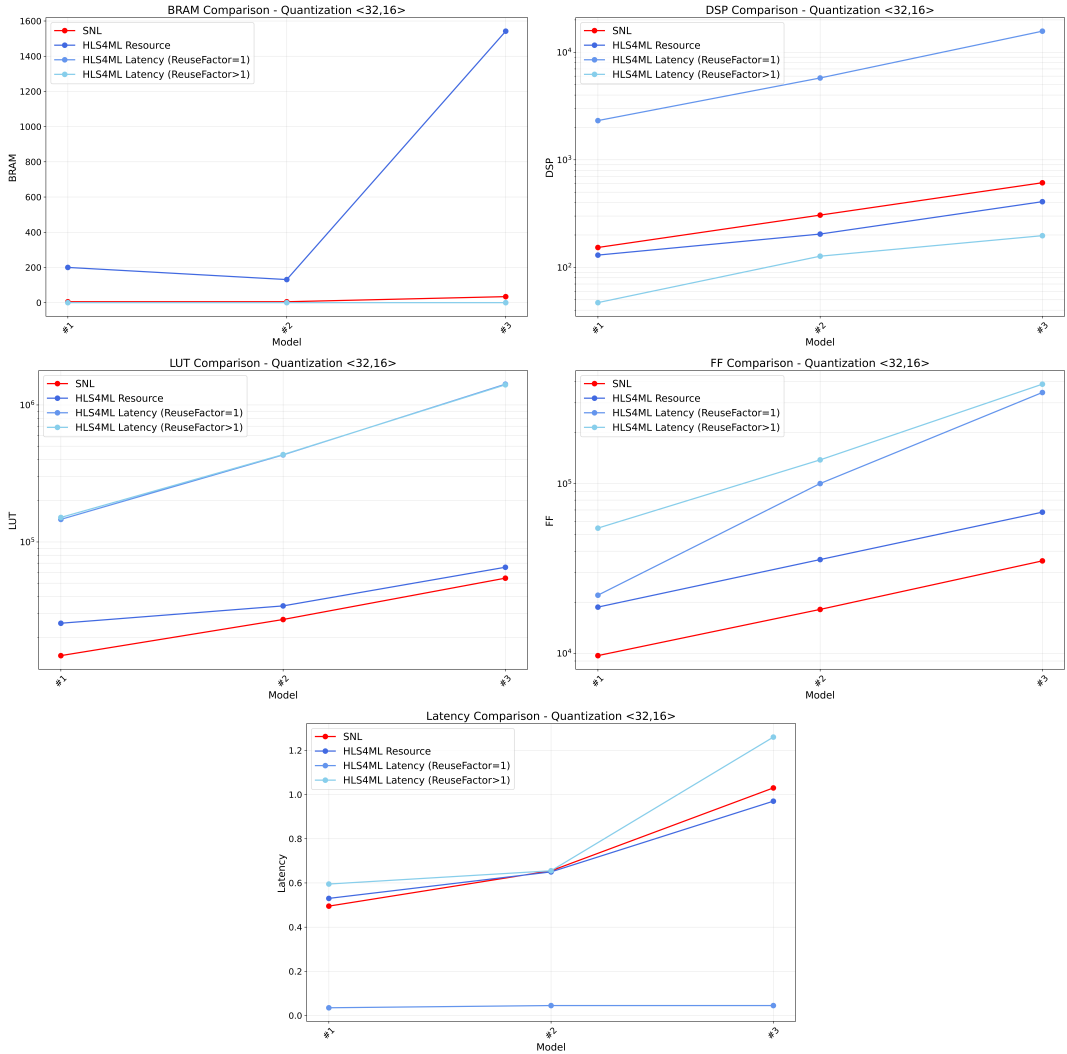

    \centering
    \includegraphics[width=0.45\linewidth]{Images/BRAM_32,16_size32_64_128.png}
    \includegraphics[width=0.45\linewidth]{Images/DSP_32,16_size32_64_128.png}
    \includegraphics[width=0.45\linewidth]{Images/LUT_32,16_size32_64_128.png}
    \includegraphics[width=0.45\linewidth]{Images/FF_32,16_size32_64_128.png}
    \includegraphics[width=0.45\linewidth]{Images/Latency_32,16_size32_64_128.png}
    \caption{Plots comparing the usage of BRAM, DSPs, LUTs, FFs, and latency for the 3 benchmark models with \texttt{ap\_fixed<16,8>} quantization.
    \label{fig:plots_168}} 
\end{figure}

Table~\ref{tab:results3216} and Table~\ref{tab:results168} show summaries of all resulting resources and latency for SNL and three \hls~synthesis strategies for the \texttt{ap\_fixed<32,16>} and \texttt{ap\_fixed<16,8>} quantization scenarios, respectively. 
Upon synthesis, some strategies resulted in implementations whose required resources exceeded that of the hardware target, which are marked in red. 
Figures~\ref{fig:plots_3216} and ~\ref{fig:plots_168} make comparisons of the resulting synthesis across all three benchmark model points for quantization to \texttt{ap\_fixed<32,16>} and  \texttt{ap\_fixed<16,8>}, respectively. 

For approximately similar latencies, the models synthesized with SNL indicate a lower FF and LUT resource usage compared to the \hls~strategies. 
The approximate linear scaling of resources with the model size indicates that this low resource usage is especially beneficial for the synthesis of very large models for future high data rate experimental applications. 
The latency-optimized \hls~implementations are able to achieve a lower overall latency than the SNL models, indicating the current outperformance of \hls~for applications with very stringent latency constraints. 
These comparative conclusions pertain only to the analysis done here and should be generalized with caution.

\begin{table}[htbp]
    \centering
    \caption{Comparison of SNL and \hls~for the three benchmark VAE trigger models with \texttt{ap\_fixed<32,16>} quantization, including BRAM, DSPs, FFs, LUTs, latency, reuse factor, and synthesis strategy. The synthesis results are for a hardware target with upper resource limits of BRAM=4320, DSP=6840, FF=2364480, and LUT=1182240. Resource requirements that exceed the available resources are marked in red. \label{tab:results3216}}
    \begin{subtable}{\textwidth}
        \centering
        \caption{Model 1 (Nodes/layer: 32/16/3)}
        \begin{tabular}{l|ccccc}
            \toprule
            \textbf{Implementation} & \textbf{BRAM} & \textbf{DSPs} & \textbf{FFs} & \textbf{LUTs} & \textbf{Latency [$\mu$s]} \\
            \midrule
            SNL & 5   & 153   & 9680  & 14795 & 0.495      \\
            \hls (Resource, matching latency) & 200 & 130  & 18707  & 25498 & 0.53-0.545 \\
            \hls (Latency, matching latency)& 0   & 47   & 54687 & 150932 & 0.595       \\
            \hls (Latency, optimized latency) & 0   & 2315 & 21988  & 146057 & 0.035      \\
            \bottomrule
        \end{tabular}
    \end{subtable}

    \bigskip
    
    \begin{subtable}{\textwidth}
        \centering
        \caption{Model 2 (Nodes/layer: 64/32/6)}
        \begin{tabular}{l|ccccc}
            \toprule
            \textbf{Implementation} & \textbf{BRAM} & \textbf{DSPs} & \textbf{FFs} & \textbf{LUTs} & \textbf{Latency [$\mu$s]} \\
            \midrule
            SNL & 5   & 306   & 18129  & 27152  & 0.665      \\
            \hls (Resource, matching latency) & 131 & 204  & 35677  & 34124 & 0.650 \\
            \hls (Latency, matching latency)& 0 & 127  & 138615  & 434187 & 0.595 \\
            \hls (Latency, optimized latency) & 0   & 5764 & 100198  & 431785 & 0.045      \\
            \bottomrule
        \end{tabular}
    \end{subtable}

    \bigskip    
    
    \begin{subtable}{\textwidth}
        \centering
        \caption{Model 3 (Nodes/layer: 128/64/12)}
        \begin{tabular}{l|ccccc}
            \toprule
            \textbf{Implementation} & \textbf{BRAM} & \textbf{DSPs} & \textbf{FFs} & \textbf{LUTs} & \textbf{Latency [$\mu$s]} \\
            \midrule
            SNL & 34   & 612   & 35049  & 54395  & 1.03      \\
            \hls (Resource, matching latency) & 1542 & 408  & 67922  & 65352 & 0.97 \\
            \hls (Latency, matching latency)& 0 & 197  & 386182  & \tcr{1403028} & 1.26 \\
            \hls (Latency, optimized latency) & 0   & \tcr{15696} & 344920  & \tcr{1417209} & 0.045      \\
            \bottomrule
        \end{tabular}
    \end{subtable}
\end{table}

\begin{table}[htbp]
    \centering
    \caption{Comparison of SNL and \hls~for the three benchmark VAE trigger models with \texttt{ap\_fixed<16,8>} quantization, including BRAM, DSPs, FFs, LUTs, latency, reuse factor, and synthesis strategy. The synthesis results are for a hardware target with upper resource limits of BRAM=4320, DSP=6840, FF=2364480, and LUT=1182240.}\label{tab:results168}
    \begin{subtable}{\textwidth}
        \centering
        \caption{Model 1 (Nodes/layer: 32/16/3)}
        \begin{tabular}{l|ccccc}
            \toprule
            \textbf{Implementation} & \textbf{BRAM} & \textbf{DSPs} & \textbf{FFs} & \textbf{LUTs} & \textbf{Latency [$\mu$s]} \\
            \midrule
            SNL &3   & 51   & 4888  & 8055  & 0.505      \\
            \hls (Resource, matching latency) & 104 & 113  & 9965  & 19132 & 0.52-0.535 \\
            \hls (Latency, matching latency)& 0   & 38   & 26846 & 81114 & 0.47       \\
            \hls (Latency, optimized latency) & 0   & 1024 & 6375  & 77567 & 0.025      \\
            \bottomrule
        \end{tabular}
    \end{subtable}

    \bigskip
    
    \begin{subtable}{\textwidth}
        \centering
        \caption{Model 2 (Nodes/layer: 64/32/6)}
        \begin{tabular}{l|ccccc}
            \toprule
            \textbf{Implementation} & \textbf{BRAM} & \textbf{DSPs} & \textbf{FFs} & \textbf{LUTs} & \textbf{Latency [$\mu$s]} \\
            \midrule
            SNL & 7   & 102   & 8436  & 13649  & 0.665      \\
            \hls (Resource, matching latency) & 451 & 262  & 19673  & 33292 & 0.450 \\
            \hls (Latency, matching latency)& 0 & 92  & 60139  & 187765 & 0.495 \\
            \hls (Latency, optimized latency) & 0   & 2399 & 26542  & 181672 & 0.035      \\
            \bottomrule
        \end{tabular}
    \end{subtable}

    \bigskip    
    
    \begin{subtable}{\textwidth}
        \centering
        \caption{Model 3 (Nodes/layer: 128/64/12)}
        \begin{tabular}{l|ccccc}
            \toprule
            \textbf{Implementation} & \textbf{BRAM} & \textbf{DSPs} & \textbf{FFs} & \textbf{LUTs} & \textbf{Latency [$\mu$s]} \\
            \midrule
            SNL & 17   & 204   & 15616  & 26169  & 1.04      \\
            \hls (Resource, matching latency) & 774 & 268  & 33609  & 45499 & 0.96 \\
            \hls (Latency, matching latency)& 0 & 127  & 163211  & 506962 & 0.74 \\
            \hls (Latency, optimized latency) & 0   & 5509 & 72717  & 502938 & 0.035      \\
            \bottomrule
        \end{tabular}
    \end{subtable}
\end{table}


\clearpage

This analysis serves as a basic proof-of-concept, and opens up a broader set of studies to pursue SNL applications to scientific experiments. 
The relative maturity, broad support, and clear documentation of \hls~indicate its clear and unique utility in the field. 
Methods such SNL will likely continue to require more engineering input and thus cannot present a comprehensive alternative, but can nonetheless further progress in the area of large-scale ML for FPGA-based physics experimentation. 
Further development of SNL is ongoing and will seek to provide more latency-based optimization methodologies to meet the \hls~performance.  

Several additional challenges must be considered when looking towards synthesis options for large-scale real-time ML in physics experiments. 
Inherent design differences between synthesis frameworks, such as parallel vs. streaming data flow, motivates careful consideration when making direct comparisons. 
The SNL implementations can only be effective in a streaming scenario, requiring considering of the layer-to-layer arithmetic with varying \texttt{ap\_fixed} precision and the ability of a hardware implementation to match the appropriate expected arrival for a given data acquisition.
Preserving numerical accuracy is also an essential test for any synthesis strategy, and subsequent works will investigate this along with resource usage. 
Future studies aim to provide robust, efficient, and user-friendly synthesis strategies to fully harness the physics potential achievable with real-time FPGA-based ML in future scientific facilities.

\section{Conclusions}
\label{sec:conclusions}


Results are presented comparing hardware synthesis frameworks, namely SNL and \hls, for an example collider trigger application involving real-time ML-based anomaly detection. 
In this context, \hls~demonstrates highly performant latency-optimized implementations, making it an excellent tool for ultra-fast applications. 
For syntheses with comparable latency, SNL delivers more resource-efficient implementations, particularly with respect to LUT and FF utilization.
This result indicates the potential for SNL to enable future deployment of larger networks on resource-constrained FPGAs. 

It is important to note that the overall system latency also depends on how data is fed into the FPGA-based neural network, as the processing cannot outpace the input arrival rate. 
Therefore, efficient optimization of FPGA resources must be considered within the broader system context. This highlights the need for a comprehensive system-level study in future work, focusing on input data handling, resource allocation, and integration of these frameworks into real-time systems. 
Additionally, expanding user support, evaluating numerical accuracy, analyzing the phase space of successful synthesis, and assessing power consumption remain critical areas for future investigation.


\section*{Acknowledgments}

This work is supported by the U.S. Department of Energy under contract number DE-AC02-76SF00515.


\bibliographystyle{iopart-num}
\bibliography{main.bib}

\end{document}